\documentclass[journal]{vgtc}            
\vgtccategory{Research}
\vgtcpapertype{system}

\newcommand{\figureScenario}{ 
\begin{figure}[!t]
    \centering
    \makebox[\linewidth]{
        \includegraphics[width=\columnwidth]{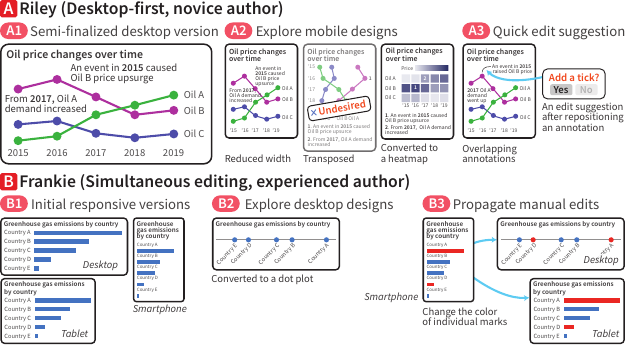}
    }
    \caption{Two proposed usage scenarios for a mixed-initiative responsive visualization design tool. \sceA{A} A novice author creating responsive designs in a desktop-first manner. \sceA{B} An experienced author editing responsive artboards simultaneously. 
    \vspace{-4mm}
    }
    \label{fig:scenario}
\end{figure}
}

\newcommand{\figureSystemOverview}{ 
\begin{figure*}[t]
    \centering
    \makebox[\linewidth]{
        \includegraphics[width=\textwidth]{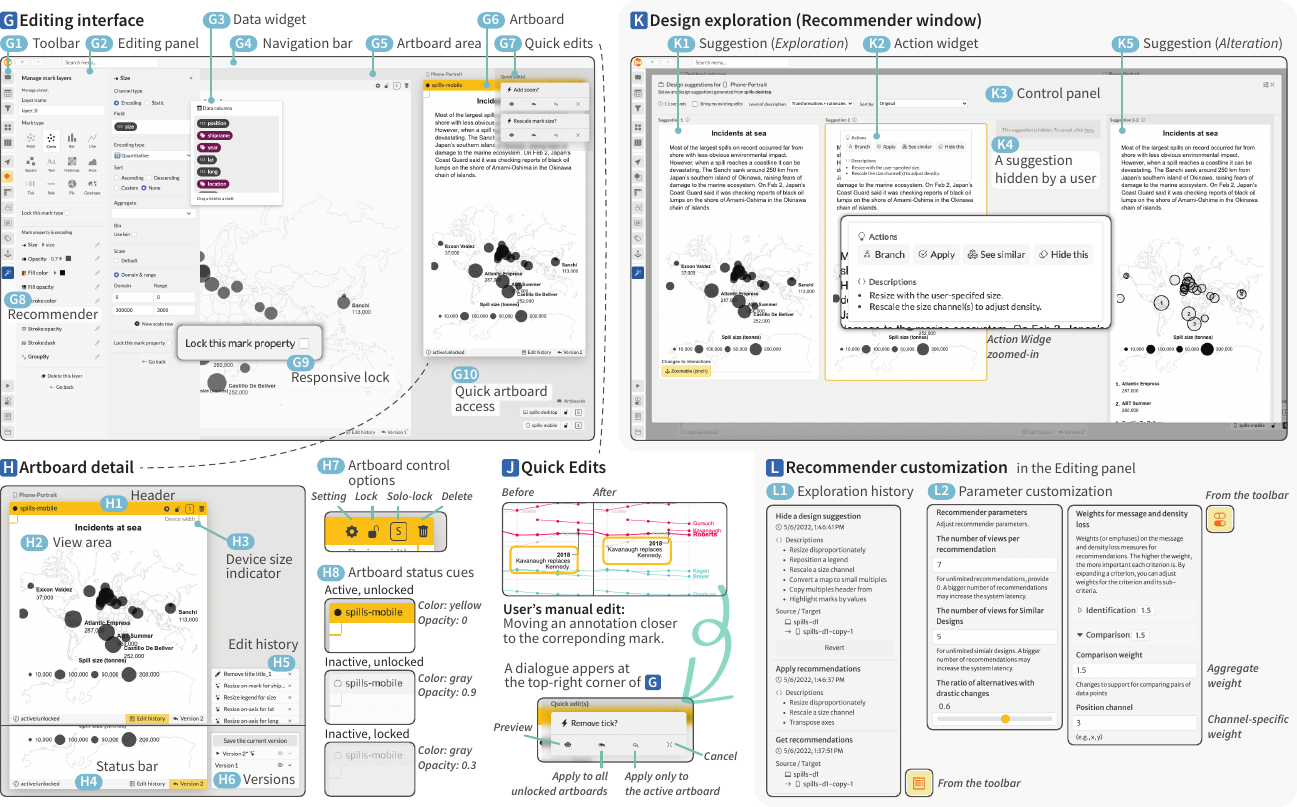}
    }
    \caption{An overview of Dupo's interface.
    \fiAFig{G} 
    By selecting the \menu{mark} menu (\makebox{\includegraphics[height=7pt]{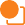}}) from the \menu{toolbar} \fiBFig{G1}, the user can edit marks (\eg~mark type and encodings) in the \menu{editing panel}~\fiBFig{G2}. To set an encoding channel, the user drags a data field from the \menu{data widget}~\fiBFig{G3} and drops it on the desired field.
    In the \menu{navigation bar} \fiBFig{G4}, the user can undo or redo by clicking the arrow icons and search the system menus.
    The \menu{artboard area}~\fiBFig{G5} displays responsive \menu{artboards}~\fiBFig{G6}.
    The user can load suggestions using the \menu{recommender button}~\fiBFig{G8}, and set \menu{responsive locks} on elements that the recommender must keep~\fiBFig{G9}.
    At the bottom right corner, the user can \menu{quickly access particular artboards}~\fiBFig{G10}. 
    \fiAFig{H} Each artboard consists of \menu{header}~\fiBFig{H1}, \menu{view area}~\fiBFig{H2}, and \menu{status bar}~\fiBFig{H4}.
    The \menu{header} provides options to manage its \menu{status} and \menu{settings} \fiBFig{H7} and indicates the \menu{activeness} and \menu{lock} status~\fiBFig{H8}. 
    In the \menu{view area}, the \menu{device size indicator}~\fiBFig{H3} helps the user to check if the content overflows the intended space.
    From the \menu{status bar}, the user can access the \menu{edit history}~\fiBFig{H5} and \menu{versions}~\fiBFig{H6}.
    \fiAFig{J}~Some manual edits by the user prompt Dupo to recommend associated \menu{quick edits}, appearing on the top right corner of the \menu{editing interface}~\fiBFig{G7}.
    \fiAFig{K}~The user has requested \menu{design suggestions}~\fiBFig{K1} generated for a smartphone version. 
    From the \menu{action widget}~\fiBFig{K2}, the user can \menu{branch} each suggestion as a new artboard, \menu{apply} it to the current artboard, request further \pipe{Alteration} suggestions, \menu{hide} a suggestion~\fiBFig{K4}, and read the rationales for each suggestion. 
    The \menu{control panel}~\fiBFig{K3} allows users to \menu{apply their custom edits} to the suggestions and toggle the depth of descriptions in the \menu{action widget}.
    \fiAFig{L}~The user can review the \menu{history}~(\makebox{\includegraphics[height=7pt]{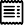}}) of their use of the recommender~\fiBFig{L1} and customize the recommender parameters~\fiBFig{L2}~(\makebox{\includegraphics[height=7pt]{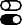}}) from the \menu{editing panel}~\fiBFig{G2}.
    \vspace{-2mm}
    }
    \label{fig:system_overview}
\end{figure*}
}

\newcommand{\figureWalkThrough}{ 
\begin{figure}[t]
    \centering
    \makebox[\linewidth]{
        \includegraphics[width=\columnwidth]{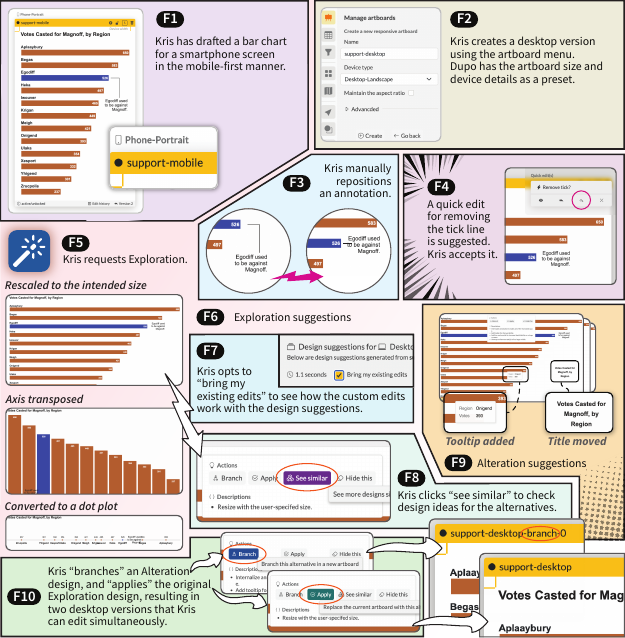}
    }
    \caption{A walkthrough example for creating a responsive visualization using Dupo in a mobile-first manner. Kris starts by drafting an initial bar chart for a mobile phone, and then uses Dupo's three recommendation pipelines to refine the design. A video walkthrough of this example is included in the Supplementary Material.
    \vspace{0mm}}
    \label{fig:walkthrough}
\end{figure}
}

\newcommand{\figureRecommenderOverview}{ 
\begin{figure*}[t]
    \centering
    \makebox[\linewidth]{
        \includegraphics[width=\textwidth]{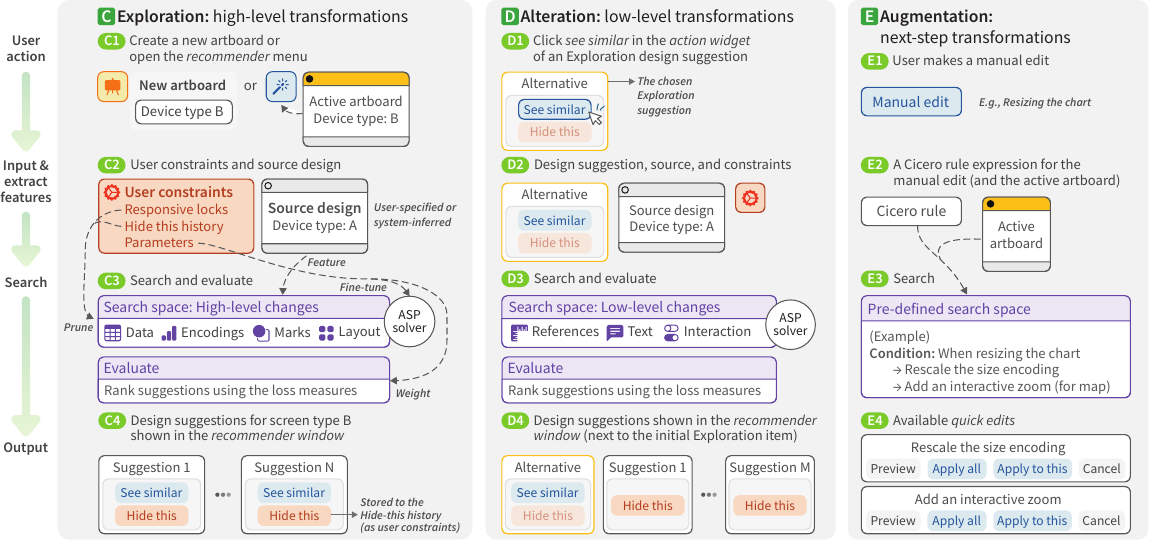}
    }
    \caption{Dupo's three recommendation pipelines: \pipe{Exploration}, \pipe{Alteration}, and \pipe{Augmentation}. Initiated by a user action, each pipeline gathers the input data, runs a search, and returns the output suggestions.
    \recAFig{C}~The user can run the \pipe{Exploration} pipeline to see high-level transformations (\eg~to data, marks, encodings, and layout) by pressing the \menu{recommender button} from the \menu{toolbar} or creating a new artboard \recBFig{C1}. Then, Dupo takes the source design and user constraints as input \recBFig{C2}. Dupo generates alternatives with high-level changes and ranks them \recBFig{C3}. These design alternatives are presented in the \menu{recommender window} \recBFig{C4}.
    If the user clicks the \menu{hide this} button of an alternative, Dupo stores the alternative design in the \menu{hide this history} as a user constraint for next time the user runs the \pipe{Exploration} pipeline.
    \recAFig{D} The user can run the \pipe{Alteration} pipeline to see low-level transformations (\eg~to references, text, interactions) by selecting \menu{see similar} for an alternative design \recBFig{D1}. Dupo takes the alternative of interest, the user constraints, and the source design as input \recBFig{D2}. Dupo then populates alternatives for low-level changes and evaluates them~\recBFig{D3}. The \pipe{Alteration} suggestions then appear next to the initial \pipe{Exploration} design \recBFig{D4}. 
    \recAFig{E} The \pipe{Augmentation} pipeline is initiated after a user makes a manual edit \recBFig{E1}, at which point Dupo takes its Cicero rule as input \recBFig{E2} and searches a pre-defined search space \recBFig{E3} to suggest \menu{quick edits} that are commonly applied with the manual edit \recBFig{E4}.
    \vspace{-3mm}
    }
    \label{fig:recommender_overview}
\end{figure*}
}

\newcommand{\figureStudyOutcome}{ 
\begin{figure*}[!t]
    \centering
    \makebox[\linewidth]{
        \includegraphics[width=\textwidth]{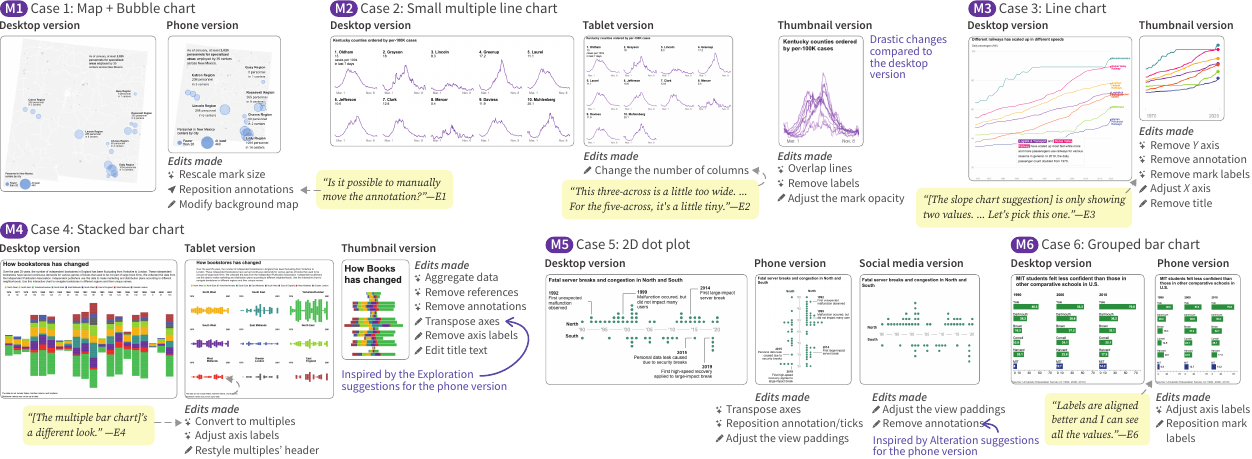}
    }
    \caption{Selective design outcomes from the user study with six expert responsive visualization authors. Participants provided their own visualizations for use in the study. For anonymization purposes, we replaced participants' data sets while maintaining the cardinality and data type, edited the text content but preserved annotation length, and altered the color schemes. A star icon (\makebox{\includegraphics[height=7pt]{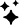}}) indicates an edit from the recommendations, a pencil icon (\makebox{\includegraphics[height=7pt]{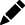}}) denotes a manual edit made by a participant, and a cursor icon  (\makebox{\includegraphics[height=7pt]{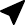}}) represents a manual edit using direct manipulation.
    \vspace{-3mm}
    }
    \label{fig:study_outcome}
\end{figure*}
}

\usepackage{xargs} 
\usepackage{soul}  
\usepackage{color} 
\usepackage{xspace}
\usepackage{listings}
\usepackage{tcolorbox}

\newcommand{\ie}{{i.e.,}\xspace}
\newcommand{\eg}{{e.g.,}\xspace}
\newcommand{\cf}{{c.f.}\xspace}
\newcommand{\ea}{{et~al.}\xspace}

\newcommand{\etc}{{etc.}\xspace}
\newcommand{\bstart}[1]{\vspace{1mm} \noindent{\textbf{#1:}}}
\newcommand{\bstartf}[1]{\noindent{\textbf{#1:}}}
\newcommand{\bpstart}[1]{\vspace{1mm} \noindent{\textbf{#1.}}}

\definecolor{activegold}{RGB}{255,193,61}

\definecolor{lightorange}{rgb}{1,0.8,0.4}
\definecolor{lightorange}{RGB}{230, 170, 50}
\definecolor{lightgreen}{RGB}{121,210,121}
\definecolor{lightteal}{RGB}{121,199,210}
\definecolor{lightblue}{RGB}{100,212,239}
\definecolor{lightpurple}{RGB}{153,102,255}
\definecolor{lightred}{RGB}{245, 132, 120}
\definecolor{red}{RGB}{178,34,34}
\definecolor{gray}{RGB}{166,166,166}
\definecolor{indexBlue}{cmyk}{0.9,0.8,0,0}
\definecolor{indexGreen}{cmyk}{0.8,0.2,0.8,0.55}
\definecolor{deepblue}{cmyk}{0.9,0.75,0,0.5}
\definecolor{deepred}{cmyk}{0,0.75,0.75,0.4}
\definecolor{pink}{RGB}{214,114,0}
\definecolor{NavyBlue}{rgb}{0.0, 0.0, 0.5}

\definecolor{sceCoA}{cmyk}{0.05, 1,    1,    0}
\definecolor{sceCoB}{cmyk}{0,    0.72, 0.49, 0}
\definecolor{intCoA}{cmyk}{0.89, 0.58, 0.1,  0}
\definecolor{intCoB}{cmyk}{0.54, 0.14, 0.12, 0}
\definecolor{recCoA}{cmyk}{0.79, 0.11, 1,    0}
\definecolor{recCoB}{cmyk}{0.52, 0,    1,    0}
\definecolor{logCoA}{cmyk}{0.02, 0.71, 1,    0}
\definecolor{logCoB}{cmyk}{0.71, 0.91, 0,    0}

\newcommand{\menu}[1]{\textsl{#1}}
\newcommand{\menuw}[1]{\emph{#1}}
\newcommand{\pipe}[1]{\emph{#1}}

\newcounter{gIndex}
\newcommand{\consideration}[2]{\refstepcounter{gIndex}\label{#2}\textbf{#1 \textcolor{indexBlue}{(DC\thegIndex)}}}
\newcommand{\gRef}[1]{\textcolor{indexBlue}{\textbf{DC\ref{#1}}}}

\newtcbox{\sceA}{nobeforeafter,tcbox raise base, 
boxrule=0mm,top=0mm,bottom=0mm,right=0mm,left=0mm,arc=0.3mm,boxsep=0.5mm,
fontupper=\sffamily\small,
colframe=sceCoA,coltext=white,colback=sceCoA}
\newtcbox{\sceB}{nobeforeafter,tcbox raise base, 
boxrule=0mm,top=0mm,bottom=0mm,right=0mm,left=0mm,arc=1.2mm,boxsep=0.5mm,
fontupper=\sffamily\small\bfseries,
colframe=sceCoB,coltext=white,colback=sceCoB}

\newtcbox{\recA}{nobeforeafter,tcbox raise base, 
boxrule=0mm,top=0mm,bottom=0mm,right=0mm,left=0mm,arc=0.3mm,boxsep=0.5mm,
fontupper=\sffamily\small,
colframe=recCoA,coltext=white,colback=recCoA}
\newtcbox{\recB}{nobeforeafter,tcbox raise base, 
boxrule=0mm,top=0mm,bottom=0mm,right=0mm,left=0mm,arc=1.2mm,boxsep=0.5mm,
fontupper=\sffamily\small,
colframe=recCoB,coltext=white,colback=recCoB}

\newtcbox{\recAFig}{nobeforeafter,tcbox raise base, 
boxrule=0mm,top=0mm,bottom=0mm,right=0mm,left=0mm,arc=0.3mm,boxsep=0.5mm,
fontupper=\sffamily\tiny,
colframe=recCoA,coltext=white,colback=recCoA}
\newtcbox{\recBFig}{nobeforeafter,tcbox raise base, 
boxrule=0mm,top=0mm,bottom=0mm,right=0mm,left=0mm,arc=1.2mm,boxsep=0.5mm,
fontupper=\sffamily\tiny,
colframe=recCoB,coltext=white,colback=recCoB}

\newtcbox{\fiA}{nobeforeafter,tcbox raise base, 
boxrule=0mm,top=0mm,bottom=0mm,right=0mm,left=0mm,arc=0.3mm,boxsep=0.5mm,
fontupper=\sffamily\small,
colframe=intCoA,coltext=white,colback=intCoA}
\newtcbox{\fiB}{nobeforeafter,tcbox raise base, 
boxrule=0mm,top=0mm,bottom=0mm,right=0mm,left=0mm,arc=1.2mm,boxsep=0.5mm,
fontupper=\sffamily\small,
colframe=intCoB,coltext=white,colback=intCoB}

\newtcbox{\fiAFig}{nobeforeafter,tcbox raise base, 
boxrule=0mm,top=0mm,bottom=0mm,right=0mm,left=0mm,arc=0.3mm,boxsep=0.5mm,
fontupper=\sffamily\tiny,
colframe=intCoA,coltext=white,colback=intCoA}
\newtcbox{\fiBFig}{nobeforeafter,tcbox raise base, 
boxrule=0mm,top=0mm,bottom=0mm,right=0mm,left=0mm,arc=1.2mm,boxsep=0.5mm,
fontupper=\sffamily\tiny,
colframe=intCoB,coltext=white,colback=intCoB}

\newtcbox{\scene}{nobeforeafter,tcbox raise base, 
boxrule=0mm,top=0mm,bottom=0mm,right=0mm,left=0mm,arc=1.2mm,boxsep=0.5mm,
fontupper=\sffamily\small,
colframe=black,coltext=white,colback=black}
\newtcbox{\sceneLong}{nobeforeafter,tcbox raise base, 
boxrule=0mm,top=0mm,bottom=0mm,right=0mm,left=0mm,arc=1.2mm,boxsep=0.5mm,
fontupper=\sffamily\small,
colframe=black,coltext=white,colback=black}

\newtcbox{\logA}{nobeforeafter,tcbox raise base, 
boxrule=0mmtop=0mm,bottom=0mm,right=0mm,left=0mm,arc=0.3mm,boxsep=0.5mm,
fontupper=\sffamily\small,
colframe=logCoA,coltext=white,colback=logCoA}
\newtcbox{\logB}{nobeforeafter,tcbox raise base, 
boxrule=0mm,top=0mm,bottom=0mm,right=0mm,left=0mm,arc=1.2mm,boxsep=0.5mm,
fontupper=\sffamily\small,
colframe=logCoB,coltext=white,colback=logCoB}

\newcommand{\walkthrough}[1]{\noindent\textsc{\underline{Walkthrough:}} \textit{#1}}

\title{Dupo: A Mixed-Initiative Authoring Tool for Responsive Visualization}

\author{%
  \authororcid{Hyeok Kim}{0000-0003-4340-4470},
  \authororcid{Ryan Rossi}{0000-0001-9758-0635}, 
  \authororcid{Jessica Hullman}{0000-0001-6826-3550}, and 
  \authororcid{Jane Hoffswell}{0000-0002-9871-4575}
}

\authorfooter{
  \item
  	Hyeok Kim and Jessica Hullman are with Northwestern University.
  	E-mail: \{hyeok, jhullman\}@northwestern.edu.
  \item
  	Jane Hoffswell and Ryan Rossi are with Adobe Research.
  	E-mail: \{jhoffs, rrossi\}@adobe.com.
}

\abstract{Designing responsive visualizations for various screen types can be tedious as authors must manage multiple chart~versions across design iterations.
Automated approaches for responsive visualization must take into account the user's need for agency in exploring possible design ideas and applying customizations based on their own goals.
We design and implement Dupo, a mixed-initiative approach to creating responsive visualizations that combines the agency afforded by a manual interface with automation provided by a recommender system.
Given an initial design, users can browse automated design suggestions for a different screen type and make edits to a chosen design, thereby supporting quick prototyping and customizability.
Dupo employs a two-step recommender pipeline that first suggests significant design changes (\textit{Exploration}) followed by more subtle changes (\textit{Alteration}). 
We evaluated Dupo with six expert responsive visualization authors. 
While creating responsive versions of a source design in Dupo, participants could reason about different design suggestions without having to manually prototype them, and thus avoid prematurely fixating on a particular design.
This process led participants to create designs that they were satisfied with but which they had previously overlooked.} 


\graphicspath{{figs/}{figures/}{pictures/}{images/}{./}} 

\usepackage[font={sf,small}]{caption}

\begin{document}


\firstsection{Introduction}\label{sec:intro}

\maketitle

\noindent Creating responsive visualizations---versions of the same visualization adapted for different screen types---is essential to communicate with a broader audience.
However, authoring responsive versions can be tedious, often requiring additional design iterations.
A common approach is to (semi-) finalize a larger design (\eg~for desktop) and then transform it for smaller screens (\eg~tablets, smartphones), or vice versa.
This process often involves making substantial design transformations beyond simple resizing~\cite{hoffswell:responsive2020,kim:responsive2021}.
Authors may try to simplify this process by falling back on design strategies they have used in the past, even if they are not optimal~\cite{linsey:designfixation2010}.
Even authors who are well versed in a variety of responsive techniques typically face a time-consuming process of manually testing and evaluating designs one by one.

Consequently, some recent work proposes automated approaches to responsive design~\cite{wu:mobilevisfixer2020,kim:insight2021} or visualization retargeting~\cite{wu:visizer2013,giacomo:network2015}. Nevertheless, in many contexts authors may want to more flexibly switch between agency and automation,  such as newsrooms and public data reporting.
While the ability to automate certain edits, like repositioning legends and rescaling size encodings, may be widely useful, authors may still need to manually edit the visualization in certain ways that are hard to effectively automate, such as rewriting text annotations to be more concise.
A mixed-initiative approach provides flexibility by allowing users to accept automated recommendations from a system and/or make their own manual edits, and has been applied in a number of visualization systems~\cite{wongsuphasawat:voyager22017,lin:dziban2020,ma:ladv2021,healey:via2008} and design tools~\cite{wall:podium2018,odonovan:designscape2015,swearngin:scout2020}.

\textbf{We designed and implemented Dupo, a mixed-initiative authoring tool} where users can create responsive communicative visualizations both through custom edits and automated assistance with flexible artboard management.
The goal of Dupo is to support authors in exploring and reasoning about different responsive versions by making it quicker and easier to prototype design alternatives. 
\textbf{Dupo implements a two-step design recommendation pipeline (\ie~\emph{Exploration} and \emph{Alteration})} to support design exploration with reduced redundancy.
Users can first retrieve and explore more significant responsive design suggestions after or while creating a version of a visualization, such as changes to the layout, encoding choices, and data transformations like (dis)aggregation and filtering (\textit{Exploration}). 
For each design recommendation, users can request further suggestions with more subtle changes to axis labels, legend position, \etc (\textit{Alteration}).
Users can also manually change the responsive versions at any time while using Dupo.
The graphical user interface of Dupo supports flexible edit propagation not only across multiple artboards for manual edits~\cite{hoffswell:responsive2020} but also between users' custom edits and the system's automated suggestions.
 
\textbf{We evaluated Dupo with six experienced responsive visualization authors who brought their own prior visualization designs} to our study in order to explore new responsive versions using Dupo.
Our participants overall believed Dupo could benefit their day-to-day responsive visualization tasks by enabling rapid and high-fidelity prototyping of possible responsive design alternatives. 
Participants were able to seamlessly make custom edits to the automated designs, demonstrating that Dupo can support progressive mixed-initiative authoring. 
We observed that Dupo supported participants in exploring different suggestions and reasoning about them as inspiration for future designs.
We finally discuss research opportunities to apply our mixed-initiative approach to other responsive visualization methods, such as template-, programming-, and graphic-based creation, and outline next steps for future software for responsive visualization within adaptive systems.

\section{Related Work}
Our work is motivated by prior work on designing responsive visualizations and mixed-initiative tools for visualization authoring.

\subsection{Responsive Visualization Authoring}
Responsive visualization design refers to creating multiple versions of a visualization to fit to different screen sizes and device types~\cite{andrews:responsive2018,hoffswell:responsive2020,kim:responsive2021} (\eg~a laptop with a keyboard and track pad; a smartphone with a touch interface).
Prior work identifies \textit{programmatic difficulties}, \textit{artboard management}, \textit{design exploration}, and \textit{changes to takeaways} as some of the major challenges in adapting visualizations for different screens.

While creating responsive visualizations demands both cross-device design and development expertise, earlier work mainly focused on programmatic techniques using D3.js~\cite{bostock:d32011}.
This programmatic support was important particularly when there was limited tooling for responsive visualization.
For example, Hinderman~\cite{hinderman:responsive2015}, K\"{o}rner~\cite{korner:responsive2016}, and Andrews~\cite{andrews:responsive2018} demonstrate the technical feasibility of creating responsive visualizations using D3.js.
The JavaScript library R3S.js~\cite{leclaire:r3sjs2015} provides direct programming interfaces for responsive visualization using~D3.js.

Authors who use GUI-based authoring tools (\eg~Adobe Illustrator with ai2html~\cite{ai2html}, Datawrapper~\cite{datawrapper}, or Flourish~\cite{flourish}) need to manage multiple \textit{artboards} (drawing or design areas) for different device types.
Authors often have a set of pre-defined screen breakpoints for responsive versions (commonly including desktops, tablets, and smartphones for the Web environment) to avoid unpredictable errors arising due to dynamic resizing on different devices~\cite{hoffswell:responsive2020,andrews:responsive2018}.
An easy approach to handling these multiple artboards is to simply finalize an initial design for a certain screen type, and only then create other responsive versions that exhibit a small set of transformations, such as resizing~\cite{hoffswell:responsive2020,kim:responsive2021}.
However, this sequential approach makes it burdensome to run design iterations on non-initial versions of the visualization, even though they are not necessarily less important to the overall success of the design. 
Instead, Hoffswell~\ea~\cite{hoffswell:responsive2020} suggest transferring edits from one artboard to the other artboards and flexibly toggling such edit transfers in order to support simultaneous design iterations on multiple artboards.
Commercial tools like ZingCharts~\cite{zingchart} and Datawrapper~\cite{datawrapper} allow authors to specify conditional settings (\eg~altering axis positions for smartphone screens) for a bundle of templates for different screen types.

Ideally, exploring design alternatives can help an author identify better ideas~\cite{kolodner:casebased1993,kolodner:improviser1996,linsey:designfixation2010}, yet manually drafting and managing more than a few design alternatives is often time-consuming.
Thus, computational (semi-) automation is often recommended~\cite{kolodner:casebased1993}, and prior work has proposed automation approaches to responsive visualization~\cite{kim:insight2021,wu:mobilevisfixer2020,wu:autolayout2021}.
Business intelligence tools like Power BI~\cite{powerbi} and Tableau~\cite{tableau} offer presets to make it easier to create day-to-day responsive visualizations for data analytics. 
MobileVisFixer~\cite{wu:mobilevisfixer2020} uses machine learning (ML) techniques to retarget non-responsively created desktop visualizations for mobile screens.
$LQ^2$~\cite{wu:autolayout2021} introduces an ML-based (pairwise) ranking model for visualization layout given a chart size (\eg~a larger number of bars or thicker bars for a wider chart size).
Kim~\ea~\cite{kim:insight2021} propose a recommendation pipeline for responsive visualizations using Answer Set Programming (ASP) that expresses a search problem in terms of facts, rules, and constraints~\cite{gebser:asp2011}.
As a means of representing such automated approaches, Kim~\ea~\cite{kim:cicero2022} present Cicero, a declarative grammar for specifying responsive transformations.

When exploring automated design alternatives, authors still need to reason about how effectively the responsive versions maintain high-level design goals such as intended takeaways for viewers~\cite{kim:responsive2021,wu:visizer2013,giacomo:network2015,rosenbaum:progressive2012,elmqvist2010:hierarchical2010}.
For example, resizing a chart can change the aspect ratio, impacting the implied visual trend or effect size, and changing the visual encodings to adjust graphical density (\eg~changing a complex line chart to a heatmap) may change the relative discriminability of points.
Reasoning about changes to a visualization ``message'' or set of key takeaways can be difficult because the same responsive design transformation strategy may result in different levels of message preservation depending on the underlying data distributions (\eg~a~relatively uniform distribution is less impacted by changes to chart size). 
Kim~\ea~\cite{kim:insight2021} propose a set of measures that approximate changes to visualization insights between responsive versions in terms of data identifiability, comparison discriminability, and trend recognition.
ViSizer~\cite{wu:visizer2013} and Di~Giacomo~\ea~\cite{giacomo:network2015} suggest automated resizing algorithms by preserving important visual features using a \mbox{significance grid}.

While prior work individually addresses various challenges in authoring responsive visualizations,
translating those techniques into an authoring system is an important step to actually support users.
For example, prior automated approaches produce one final outcome~\cite{wu:mobilevisfixer2020} or too many outcomes with redundancy~\cite{kim:insight2021}, limiting users' ability to explore designs.
Informed by these works, we implement three types of automated recommender pipelines (\textit{Exploration}, \textit{Alteration}, and \textit{Augmentation}) and incorporate them with flexible customization within a unified end-user system for authoring responsive visualizations.

\subsection{Mixed-initiative Visualization Authoring}
Prior work on visualization tools (\eg~for ML-driven analytics~\cite{wall:podium2018}, exploratory data analysis~\cite{wongsuphasawat:voyager2016,wongsuphasawat:voyager22017,lin:dziban2020,cook:ade2015},  business intelligence~\cite{tableau}, dashboard design~\cite{ma:ladv2021,qu2017keeping}, chart design~\cite{healey:via2008}, and infographics~\cite{cui:infographics2022}) has applied a mixed-initiative approach where users can make manual changes or apply system suggestions.
One benefit of mixed-initiative approaches for authoring is that it expedites design iterations by reducing the human effort involved in prototyping.
However, prior recommenders for responsive visualization design~\cite{wu:mobilevisfixer2020,kim:insight2021} are intended to be one-shot recommendations, lacking considerations for progressive design processes where users make manual edits and request automation iteratively~\cite{lin:dziban2020,ma:ladv2021,wongsuphasawat:voyager2016,healey:via2008}.
Our work outlines a recommender pipeline suited for mixed-initiative authoring of responsive visualizations.

\section{Usage Scenarios}\label{sec:usage}

To motivate our design considerations for mixed-initiative responsive visualization authoring tools, we propose two usage scenarios in which a visualization author uses a mixed-initiative tool, based on prior formative findings~\cite{hoffswell:responsive2020,kim:responsive2021} and recommender approaches~\cite{kim:insight2021,wu:mobilevisfixer2020}.

\figureScenario{}

\figureRecommenderOverview{}

\bstart{Desktop-first, novice author}
Riley (\autoref{fig:scenario}~\sceA{A}) is a novice visualization author with limited experience in creating responsive designs.
Riley uses a desktop device during the development process, and thus decides to start by designing a desktop visualization.
After roughly finishing the desktop design of a line chart about oil price changes over time~\sceB{A1}, Riley explores mobile versions suggested by the mixed-initiative tool~\sceB{A2}.
When examining the recommendations, Riley notices that some of the suggestions violate a design guideline in Riley's organization that discourages putting a temporal field on a vertical axis in a line chart.
To filter that case out, Riley marks it as ``undesired,'' so that the system does not recommend similar cases next time Riley retrieves design recommendations in this scenario.
Riley thinks two versions, one with the reduced width and the other converted to a heatmap, seem useful. Riley selects both of them to discuss with their team.
Riley realizes that some annotations are overlapping for these designs and decides to fix the position and width of annotations to avoid overlap and make them more readable~\sceB{A3}.
By doing so, one of the annotations is repositioned further away from the data point that it is indicating, so the tool suggests adding a tick between that distant annotation and the corresponding data mark.
Riley accepts the quick edit suggestion.

\bstart{Simultaneous editing, experienced author}
Frankie (\autoref{fig:scenario}~\sceA{B}) is an experienced visualization designer.
Frankie prefers to create responsive designs simultaneously by managing multiple artboards together, which makes inspecting and comparing takeaways between versions easier.
Frankie starts with artboards for desktop, tablet, and mobile, which are mandated screen breakpoints in Frankie's organization.
After creating a horizontal bar chart about greenhouse gas emissions by country~\sceB{B1}, Frankie realizes that the desktop version looks too wide and does not use the large screen space efficiently.
Thus, Frankie retrieves some design recommendations using the mixed-initiative tool for the desktop version to reduce the time it takes to prototype different ideas~\sceB{B2}.
Frankie finds a one-dimensional dot plot that spreads the quantities (previously encoded by the bar length) along the same horizontal axis, which better uses the landscape aspect ratio of desktop devices without distorting the original takeaway about the rank of those countries.
Frankie selects the dot plot version to further evaluate alongside the original bar chart.
Frankie now wants to highlight two countries to indicate some important points~\sceB{B3}. 
Thus, Frankie changes the color of the marks for those countries (in the mobile version) to be contrasting. This edit is propagated to the desktop and tablet versions.

\section{Design Considerations}\label{sec:guidelines}
\noindent To motivate the design of our mixed-initiative responsive visualization authoring tool, Dupo, we derive the following four design considerations inspired by prior work and our usage scenarios (\autoref{sec:usage}).

\bstart{\consideration{Support design exploration}{dg:exploration}}
A benefit of automation is quickly enumerating and ranking recommendations from a large design space, so that users can efficiently explore many different options and avoid design fixation.
To not overwhelm users with subtle differences, a recommender should minimize redundancy among its suggestions~\cite{wongsuphasawat:voyager22017,odonovan:designscape2015}.
To support reasoning about the critical trade-off between preserving visualization messages and maintaining graphical density~\cite{kim:responsive2021,kim:insight2021}, a recommender and its interface should employ methods capable of evaluating how well designs preserve important patterns or insights.

\bstart{\consideration{Support user control}{dg:control}}
Responsive visualization authors need to retain their agency by being able to control and understand the automation process~\cite{hoffswell:responsive2020}.
In particular, users should be able to fine-tune the systems' recommendation methods (\eg~conveying which suggestions are less useful to guide future recommendation strategies).
To support authors in making well-informed decisions about recommended alternatives, a mixed-initiative system should provide easy-to-interpret descriptions of what each suggestion is about and why it is suggested.

\bstart{\consideration{Support progressive authoring}{dg:progressive}}
Mixed-initiative systems should support an iterative design process that facilitates seamlessly switching between customization and automation~\cite{lin:dziban2020,ma:ladv2021,wongsuphasawat:voyager2016,wongsuphasawat:voyager2016,healey:via2008}.
To enable progressive authoring, users should be able to easily retrieve recommendations on demand, and recommended visualizations should be customizable as needed.
In addition, a progressive authoring interface should support multiple ways to revisit and compare prior~designs.

\bstart{\consideration{Support flexible artboard management}{dg:flexibility}}
Users may have different preferences in creating responsive visualizations (\eg~desktop-first, simultaneous editing).
Hoffswell~\ea~\cite{hoffswell:responsive2020} provide four design guidelines for flexible artboard management.
Users should be able to quickly preview and edit multiple artboards simultaneously as well as customize a specific artboard and propagate edits between designs.
To support progressive authoring (\gRef{dg:progressive}), such flexibility should also enable propagation between custom edits and automated design suggestions.

\section{Recommendation Pipeline}\label{sec:recommendation}

\noindent To facilitate \textbf{design exploration} (\gRef{dg:exploration}), Dupo provides design suggestions using three pipelines: \pipe{Exploration}, \pipe{Alteration}, and \pipe{Augmentation}, as shown in \autoref{fig:recommender_overview}.
For the main recommendation workflow, Dupo provides design suggestions in two steps: \pipe{Exploration} and \pipe{Alteration}.
The \pipe{Exploration} pipeline (\autoref{fig:recommender_overview} \recA{C}) suggests design alternatives with high-level changes mainly to mark types, layout, data transformations, and encodings.
The \pipe{Alteration} pipeline \recA{D} suggests design alternatives with subtle, low-level changes to text elements (\eg~annotations, titles, and labels), tooltips, references, \etc
The differentiation into these two steps aims to reduce redundancy during \textbf{design exploration}~(\gRef{dg:exploration}), and enable fine-grained refinement of preferred recommendations~(\gRef{dg:control}).

Dupo also provides a simplified \pipe{Augmentation} pipeline~\recA{E} that recommends a single responsive transformation as a possible next step in response to manual user edits. 
These recommendations (also known as \menu{quick edits}) encode commonly co-occurring design patterns from the prior responsive visualization literature~\cite{hoffswell:responsive2020,kim:responsive2021}.

\subsection{Exploration: High-level Transformations}\label{sec:exploration}

\bstartf{1. Input} The \pipe{Exploration} pipeline takes a source design and set of user constraints as input~\recB{C2} in order to provide \textbf{user control} (\gRef{dg:control}) over the outcome of the recommendation pipeline. 

\bpstart{1a. Source design}
Dupo uses the source design to determine inapplicable transformations while maintaining the underlying data characteristics (\eg~data type), aesthetic choices (\eg~encoding scales), and text contents.
For example, if the source design is a pie chart, then transformation strategies for a bar chart (\eg~transposing) are omitted from the search space unless there is a change to the mark type.
During the evaluation stage, ranking measures compare each suggested outcome with the source design in terms of how well they preserve the design information. 
The source design also helps maintain consistency, which reflects common responsive visualization authoring practices where authors apply transformations to an initial version to produce other versions, as described in the usage scenario~(\autoref{fig:scenario}~\sceA{A}).

\bpstart{1b. User constraints}
The \pipe{Exploration} pipeline leverages three types of user constraints as input: (1)~\menu{responsive locks} that provide explicit preferences about what elements cannot be updated; (2)~\emph{hidden recommendations} that provide implicit feedback about what transformations are undesirable; and (3)~\emph{recommender parameters} that provide abstract feedback on the outputs and weights applied by the recommender.
First, the user can apply two types of \menu{responsive locks}: \menu{element-locks} and \menu{position-locks}.
An \menu{element-lock} allows the user to specify elements they want to prevent from being removed and/or restyled.
For example, if the user \menu{element-locks} a legend, then Dupo ensures that all alternatives will include a legend, but still allows for repositioning the legend within the design. 
To prevent repositioning of the element, the user can specify a \menu{position-lock}.
We distinguish between an \menu{element-lock} and a \menu{position-lock} to enhance the degree of freedom for controlling the recommender given that authors do not always modify and reposition elements together~\cite{kim:responsive2021}. 
Using these \menu{responsive locks}, Dupo prunes suggestions with the locked elements from the search space.

Second, when viewing recommendations in Dupo, the user can choose to \menu{hide this} suggestion~(\ie~mark it as undesired) to implicitly update the behavior of the recommender. Dupo then records the unique transformations of that hidden suggestion (\ie~the transformations that do not appear in any other recommendations at the time) and removes them from the search space for subsequent rounds of recommendation. 

Third, the user can provide recommender parameters including the maximum number of suggestions (at each time of request), the weights for the ranking measures, and how ``drastic'' the transformations are.
Here, non-drastic transformations are those that only apply to resizing and transposing; drastic transformations include those that update encodings and layout beyond transposing.
For example, if the maximum number of suggestions is five and the drastic parameter is 0.6, then three suggestions will include drastic changes.

\bpstart{2. Search space generation}
When the \pipe{Exploration} pipeline is initiated, Dupo generates a search space~\recB{C3} that combines predefined responsive transformation techniques with the user constraints outlined above. 
Technically, the search space consists of facts describing the user input, rules encoding the responsive transformations, and hard user-specified constraints, expressed using the ASP grammar~\cite{gebser:asp2011}.
Dupo uses Clingo (an ASP solver)~\cite{gebser:clingo2014} to compute the search space for a set of suggestions.
These suggestions are translated into Cicero specifications~\cite{kim:cicero2022} so that the Cicero compiler can convert the design accordingly.

These pre-defined techniques include high-level transformation rules for mark type, layout (rows and columns; small multiples), data transformations (\eg~filtering or aggregation), and encoding channels~\recB{C3}.
When required for well-formedness, these techniques include changes to text elements (\eg~repositioning annotations when transposing the axes).
Transformation strategies for the \pipe{Exploration} search space are motivated by several goals: minimize changes between responsive designs, avoid overplotting, fit to the new aspect ratio, and maintain graphical density (\ie~the white space and overlapping area).
For each rationale, we encode transformation strategies observed in prior work~\cite{kim:responsive2021,hoffswell:responsive2020,kim:cicero2022}.
For example, Dupo suggests aggregation when the outcome device size is smaller than the source to avoid overplotting.
We document these rationales and strategies in the Supplementary Material.

Transformation strategies encoded in our search space are applied based on the device and responsive authoring direction (\eg~mobile-first, desktop-first).
Chart resizing and transposing axes, for example, are the default strategies across those conditions.
While Dupo suggests further \pipe{Exploration} suggestions, it only suggests those default transformations for tablets in a desktop-first direction unless requested by the user.
Dupo suggests moving annotations out of the chart for desktop-first whereas it suggests moving titles into the chart for mobile-first.

\bpstart{3. Ranking measures}\label{sec:ranking}
After generating design alternatives from the search space and before returning them to the user, Dupo evaluates them using several ranking measures in comparison with their source design in terms of ``message'' and density~\recB{C3}.
The ranking measures are motivated by a trade-off from prior work between preserving takeaways and adjusting density in responsive visualizations~\cite{kim:responsive2021}.
We use the task-oriented loss measures for \emph{identification}, \emph{comparison}, and \emph{trend} proposed by Kim~\ea~\cite{kim:insight2021} to approximate changes to  a visualization's support for those tasks as a proxy for ``message.''
We added a \emph{text} loss that estimates changes to the text elements (whether removed or changed), given their importance in communicative visualizations.
For density, we included an \emph{overplotting ratio} (as a proportion of the overplotted area when elements are overlapping) and an \emph{occupation ratio} (as a proportion of the non-white space out of the entire chart area), inspired by the clutter reduction method proposed by Ellis~\ea~\cite{ellis:clutter2006}.

Provided as a way to express users' different priorities, these six measures (four insights and two density) are combined into a single scalar value as a weighted sum.
By testing them on example use cases, we heuristically determined the initial weights in a way that prioritizes trend insight.
Users can also \textbf{manually fine-tune the weights} in the interface~(\gRef{dg:control}).
While we do not claim these measures as our contribution, we provide additional details in the Supplementary Material.

\subsection{Alteration: Low-level Transformations}
\label{sec:alteration}

\noindent The user can request \pipe{Alteration} suggestions for each \pipe{Exploration} suggestion (\autoref{fig:recommender_overview} \recB{D1}).
The \pipe{Alteration} pipeline suggests low-level changes to text elements (\eg~title, annotations), references, tooltips, \etc\ that make relatively subtle modifications to the design. 
Given that some \pipe{Exploration} suggestions may already include low-level changes in order to ensure well-formedness,
the \pipe{Alteration} pipeline takes the \pipe{Exploration} suggestion as input to prune duplicate transformations from the search space, in addition to leveraging the source design and user constraints~\recB{D2}.
The search space encodes a variety of the low-level changes documented by prior work~\cite{kim:responsive2021,hoffswell:responsive2020}~\recB{D3}.
For example, \pipe{Alteration} from phone to desktop suggests adding labels for a quantitative axis, internalizing a short title, and adding a tooltip.
Users can request \pipe{Alteration} multiple times for the same \pipe{Exploration} suggestion.

\subsection{Augmentation: Next-step Transformations}\label{sec:augment}
Prior work indicates that some responsive strategies are commonly applied together~\cite{kim:responsive2021}, so Dupo suggests \menu{quick edits} that recommend next-step transformations for a manual user edit.
As shown in \mbox{\autoref{fig:recommender_overview}~\recA{E}}, whenever a user makes a manual edit~\recB{E1}, Dupo passes the Cicero rule representing the edit~\recB{E2} to a pre-defined search space~\recB{E3}, and then suggests applicable \menu{quick edits} in the user interface~\recB{E4}.
To identify relevant suggestions, the \pipe{Augmentation} search space considers the direction of the edit and the intended device type.
For instance, Dupo suggests a \menu{quick edit} for fixing the tooltip position at the bottom when a tooltip is \textit{added} (direction) for a \textit{phone} version (device type).
Dupo supports a variety of \menu{quick edits} including adding or removing tick lines after repositioning annotations and moving the title text into the chart area when it is added for desktop or tablet versions, for example.
We outline the entire set of \menu{quick edit} rules in the Supplementary Material.
\figureWalkThrough{}

\figureSystemOverview{}

\vspace{-2mm}
\section{Dupo User Interface}
\label{sec:system}

\noindent Based on our design considerations (\autoref{sec:guidelines}), we implement Dupo, a mixed-initiative authoring tool for responsive visualization.
Dupo supports designing communicative visualizations with common visual encodings, annotations, and interactivity (including zoom+pan, tooltips, a brush-based context view, and an interactive legend for selecting marks on hover).
The following sections describe the key features of Dupo, as illustrated by the walkthrough in \autoref{fig:walkthrough}; we provide a user manual with a video demo detailing all of the functionality of Dupo and a walkthrough video in the Supplementary Material.

\subsection{Overview}

\walkthrough{Kris starts using Dupo by manually drafting an initial bar chart for a smartphone screen in a mobile-first manner}~\scene{F1}.

\vspace{1mm}\noindent As shown in \autoref{fig:system_overview}, Dupo has two key components: an \menu{editing interface}~\fiA{G} for manually editing the visualization through pre-defined marks (with layers) and drag-and-drop interactions, and a \menu{recommender window}~\fiA{K} to present the responsive recommendations for the current version. 
The \menu{editing interface} of Dupo consists of a \menu{toolbar}~\fiB{G1}, \menu{editing panel}~\fiB{G2}, \menu{navigation bar}~\fiB{G4}, and \menu{artboard area}~\fiB{G5}.

As the main control menu of Dupo, the \menu{toolbar} offers functionalities for manual edits (\eg~updating properties of the \menu{mark} or \menu{layout}) and system features (\eg~\menu{preview} and \menu{export}).
A user can access the menus by using the \menu{toolbar}, pressing a keyboard shortcut, or double-clicking a chart element, which opens an \menu{editing panel} specific to the chosen menu. 
As shown in \autoref{fig:system_overview}~\fiA{G}, for example, when the \menu{mark} menu~(\includegraphics[height=7pt]{figures/icons/mark-selected.pdf}) is selected, the \menu{editing panel} offers options for customizing the mark type and details (\eg~use a point on line marks) and mark property encoding channels (\eg~fill color or mark size).
Following the drag-and-drop variable assignment in popular tools like Tableau Software, the user can drag a field from the floating \menu{data widget} \fiB{G3} and drop it on the field form (or shelf). 
Dupo also enables limited direct manipulation for repositioning annotations directly from the \menu{artboard area}.

The \menu{navigation bar} offers undo and redo options and a search form for quickly accessing relevant menus.
At the bottom right corner, the user can access each artboard from the artboard list (\menu{quick artboard access})~\fiB{G10},
which scrolls to the selected artboard in the \menu{artboard area}.

\subsection{Artboard Management}
\walkthrough{Kris creates a desktop version using the artboard menu by selecting the ``Desktop-Landscape'' preset, which has a default artboard size and relevant device details already defined}~\scene{F2}.

\vspace{1mm}\noindent 
The \menu{artboard area}~(\autoref{fig:system_overview}~\fiB{G5}) displays responsive artboards, and allows a user to \textbf{see and edit multiple artboards simultaneously}~(\gRef{dg:flexibility}), inspired by Hoffswell~\ea~\cite{hoffswell:responsive2020}.
An \menu{artboard interface}~\fiA{H} consists of a \menu{header}~\fiB{H1}, \menu{view area}~\fiB{H2}, and \menu{status bar}~\fiB{H4}. 
The \menu{header} includes the name of the artboard and indicates its status~\fiB{H8}, as well as providing buttons for artboard settings (\eg~artboard size and targeted device size), lock, solo-lock, and deletion \fiB{H7}.
The \menu{view area}~\fiB{H2} of an artboard displays the responsive design.
On the top and left offsets of the \menu{view area}, \menu{device size indicators} \fiB{H3} (\ie~the yellow ticks), analogous to rulers in graphical software, help the user to check whether the content overflows the screen of the intended device.

The \menu{status bar}~\fiB{H4} of an artboard provides access to the \menu{edit history} and \menu{version list} to allow authors to revisit previous designs~(\gRef{dg:progressive}).
If a user clicks the \menu{edit history}~\fiB{H5} button, then they can see the list of manual and automated edits.
To help users distinguish the source of edits for better overall \textbf{control} (\gRef{dg:control}),
Dupo uses different icons:
a pencil icon~(\includegraphics[height=7pt]{figures/icons/pencil.pdf}) for manual edits,
a cursor icon~(\includegraphics[height=7pt]{figures/icons/selection.pdf}) for direct manipulation,
a star icon~(\includegraphics[height=7pt]{figures/icons/stars.pdf}) for edits from design suggestions, and
a lightening icon~(\includegraphics[height=7pt]{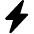}) for augmented quick edits.
A user can undo each edit by clicking the cancel button~(\includegraphics[height=7pt]{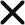}).
In the \menu{version list}~\fiB{H6}, a triangle icon~(\includegraphics[height=7pt]{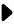}) indicates the current version, and the star icon~(\includegraphics[height=7pt]{figures/icons/stars.pdf}) denotes a version suggested by Dupo.
When a user wants to compare the current version with a previous one, they can preview a version with the eye icon~(\includegraphics[height=7pt]{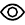}).
If the user wants to revert to a previous version, then they can ``check out'' a version with the check-mark icon~(\includegraphics[height=7pt]{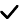}).
The user can save the current artboard at any time as a new version in the \menu{version list} to provide more \textbf{user control} over the state of the responsive versions~(\gRef{dg:control}).

Given that Dupo supports \textbf{edit propagation across multiple artboards}~(\gRef{dg:flexibility}; \cf~\cite{hoffswell:responsive2020}), it is important to enable the user to \textbf{control and see related status information}~(\gRef{dg:control}).
Each artboard maintains an \menu{activeness} status to indicate whether it is the source of changes for edit propagation (`from') and a \menu{lock} status to denote whether it is receiving propagated edits (`to').
An \menu{active} artboard (which applies to only one artboard at a time) refers to the artboard that a user is making changes to.
Each custom edit made to the active artboard is propagated only to the other \menu{unlocked} artboards.
The user can activate an unlocked artboard by clicking it, indicated by the yellow \menu{header} color and in the \menu{status bar}.
The lock button in the artboard \menu{header}~\fiB{H7} toggles and shows the lock status (locked:~\includegraphics[height=7pt]{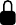}, unlocked:~\includegraphics[height=7pt]{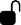}).
To edit a single artboard without edit propagation, the user can click \menu{solo-lock} to lock all the other artboards~(\includegraphics[height=7pt]{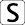}).
The \menu{status bar}~\fiB{H4}, opacity, and header color~\fiB{H8} of an artboard cues the \menu{lock} and \menu{activeness} status at a glance.

\subsection{Edit Augmentation (\menu{Quick Edits})}

\walkthrough{Kris manually moves an annotation closer to the corresponding data mark for the desktop version}~\scene{F3}\textit{. 
Dupo then suggests a \menuw{quick edit} to remove the tick line
connecting them as their relationship looks clearer}~\scene{F4}\textit{, which Kris accepts.}

\vspace{1mm}\noindent After a user makes a manual edit, Dupo suggests \menu{quick edits} that are commonly combined with the user's edit for other responsive visualization use cases~\cite{kim:responsive2021} (\pipe{Augmentation}), as shown in~\autoref{fig:system_overview}~\fiA{J}.
Applicable \menu{quick edits} appear on the top right corner of the \menu{artboard area}~\fiB{G7}, with options to preview a \menu{quick edit}, apply it to the current artboard or all the unlocked artboards, and cancel it. 

\subsection{Responsive Design Recommendations}
As described in \autoref{sec:recommendation}, Dupo distinguishes the primary recommendation workflow into a two-step approach: high-level changes to mark types, layout, data transformations, and encodings~(\pipe{Exploration}) and low-level transformations regarding references, annotations, text elements, and more~(\pipe{Alteration}).
Responsive recommendations for the current \menu{active} visualization are shown in the \menu{recommender window}~\fiA{K}.

\subsubsection{Design Exploration}

\walkthrough{To continue refining the design, Kris clicks the \menuw{recommender button} to request new \pipe{Exploration} suggestions for the desktop version} \sceneLong{F5--F6}.
\textit{Before exploring them in the \menuw{recommender window}, Kris first \menuw{applies the previous edits}} \scene{F7} \textit{to all the suggestions.}

\vspace{1mm}\noindent After (semi-) finalizing an artboard for a certain screen type (\ie~after indicating a dataset, a layout, and a mark layer), the user can explore design alternatives by creating another artboard for a different screen type or by clicking the \menu{recommender button} (\includegraphics[height=8pt]{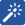} in \autoref{fig:system_overview}~\fiB{G8}) in the \menu{toolbar}.
Dupo then opens the \menu{recommender window}~\fiA{K} consisting of the \menu{control panel}~\fiB{K3} and design suggestions.
The \menu{recommender window} initially shows design suggestions for the current active artboard that consist of combinations of changes made to high-level visualization elements like visual encodings, layout, and data (\ie~\pipe{Exploration} suggestions).
The order of the suggestions implicitly encodes
Dupo's estimate of the effectiveness of the designs, as described in \autoref{sec:ranking}.
To support \textbf{progressive authoring}~(\gRef{dg:progressive}), Dupo lets the user apply the edits from the current active artboard (if applicable) to the design suggestions using the \menu{control panel}.
To do so, Dupo applies the user's manual edits (as Cicero rules) to the automated transformation rules so that those manual edits are displayed.

\subsubsection{Interaction with Design Suggestions}

\walkthrough{Kris finds an appealing recommendation for a resized bar chart, but is not completely satisfied with it. 
An \menuw{action widget} shows up as Kris hovers over the suggestion.
Kris reads the description of the design changes, and clicks \menuw{see similar} to explore other \pipe{Alteration} examples} \scene{F8}\textit{.
For example, one suggestion involves moving the title into the chart and adding a tooltip}~\scene{F9}\textit{.} 
\textit{After exploring design suggestions, Kris decides to \menuw{branch} an \pipe{Alteration} design as a separate version and \menuw{apply} the initial \pipe{Exploration} suggestion to the active artboard}~\scene{F10}\textit{.}

\vspace{1mm}\noindent When the user hovers on a suggestion, an \menu{action widget}~(\autoref{fig:system_overview}~\fiB{K2}) allows them to \menu{load similar} design suggestions, \menu{apply} the suggestion to the current artboard, \menu{branch} the suggestion as a new artboard, and \menu{hide the suggestion}.
To help users understand what transformations were applied to the design and why, the \menu{action widget} shows the list of the responsive transformations (\eg~resizing, encoding changes) applied to the suggestion.
If the user clicks the \menu{see similar} button in the \menu{action widget}, Dupo shows new alternatives~\fiB{K5} that include low-level changes to the text placement or style (\ie~\pipe{Alteration} suggestions).

The \menu{branch} and \menu{apply} options allow the user to further refine \textbf{designs they explored} (\gRef{dg:exploration}) and provide different ways to revisit previous designs to support \textbf{progressive authoring} (\gRef{dg:progressive}). 
If the user chooses to \menu{branch} a suggestion from the \menu{action widget}~\fiB{K2}, a new artboard is created with the chosen design, thereby preserving the current artboard and allowing the user to compare and iterate on multiple versions of the design.
If the user chooses to \menu{apply} a suggestion, Dupo saves the existing design in the version history~\fiB{H6} and updates the artboard with the chosen design.
The user can then make manual edits to \textbf{progressively} refine the design~(\gRef{dg:progressive}).
When the user clicks \menu{hide this}, Dupo immediately removes the suggestion~\fiB{K4} and records the recommendation as one not to suggest again.
The user can cancel the \menu{hide this} behavior immediately or revert it from the \menu{exploration history}~\fiB{L1}, in order to better \textbf{control} the recommender behavior~(\gRef{dg:control}). 

\subsection{User Controls}\label{sec:system:control}
Dupo enables the user to \textbf{control the recommender} (\gRef{dg:control}) in various ways, including applying \menu{responsive locks}, reviewing the \menu{exploration history}, adjusting \menu{recommender parameters}, and \menu{quick sorting}. 
The user can indicate visualization elements (\eg~marks or layout) that they want to prevent the recommender from changing~(\autoref{fig:system_overview}~\fiB{G9}) within the \menu{editing interface}.
The user can review their interactions with the recommender in the \menu{exploration history} menu~(\includegraphics[height=7pt]{figures/icons/exploration-history.pdf});
this menu shows the user what suggestions they have applied, branched, or hidden, and allows the user to revert \menu{hide this} decisions \fiB{L1}.
The \menu{preference} menu~(\includegraphics[height=7pt]{figures/icons/preferences.pdf}) allows the user to adjust recommender parameters~\fiB{L2}. 
As a simplified fine-tuning method, the user can choose a sorting criterion as either identification, comparison, trend recognition, text content, or graphical density from the \menu{control panel}~\fiB{K3}. 
In the \menu{control panel}, the user can also adjust the level of detail in the descriptions (\ie~transformations only or transformations with rationales).

\subsection{General System Features}
Dupo supports common GUI-based user interaction for graphical software, such as color pickers and sliders.
To preview responsive versions together, the user can use the \menu{device preview} menu (\includegraphics[height=7pt]{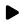}) that rescales visualizations using the pixel per inch (PPI) value of each artboard as a way to quickly verify the output of their responsive views.
Dupo lets the user \menu{export} their artboards as an HTML file with media queries~(\includegraphics[height=7pt]{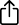}) that ensure that each artboard appears only for the specified browser size and/or aspect ratio (if provided).
In addition, Dupo offers several general system \menu{preference} options, such as switching device size indicators on and off, and changing the default artboard presets. 

\subsection{System Design Refinement}
We refined the interface based on feedback from a pilot study with~four visualization designers (see the Supplementary Material for details).
For example, Dupo originally had two ways to add a responsive artboard: using the \menu{artboard} menu to create a blank artboard and duplicating an existing artboard with a new size.
Users could specify the source design for the duplicated artboard as the original one, which was intended to enables users to specify source designs flexibly.
However, our pilot participants were confused about having multiple options for the seemingly same task. 
To consolidate these options, by default we duplicate the source design (the earliest created artboard) when the user creates a new artboard in the \menu{artboard} menu, rather than requiring a separate pipeline to produce this behavior. 
We also provide an option to change the source design for the recommendation to retain the same flexibility.
There were also originally separate buttons for high-level and low-level design suggestions; in the revised system, we moved the button for low-level suggestions to the \menu{action~widget}~(\autoref{fig:system_overview}~\fiB{K2}).

\subsection{Implementation Details}
Dupo renders visualizations using Vega-Lite~\cite{satyanarayan:vega-lite2017} with our own custom extensions for common Web-based communicative visualization techniques (\eg~text wrapping and positioning annotations relative to corresponding data marks).
To support \textbf{progressive authoring} (\gRef{dg:progressive}), it is important to streamline human edits and automated suggestions, so that users can revisit both manual and automated edits.
To do so, Dupo uses the Cicero~\cite{kim:cicero2022} grammar to express both human and automated responsive design transformations.
Each Cicero rule is composed of a \textit{specifier} (what to change; \eg~axis, layout), \textit{action} (type of change; \eg~add, modify), and \textit{option} (how to change; \eg~font size, mark type).
Whenever there are new Cicero-expressed edits, Dupo compiles it with other existing rules to update the visualization in each artboard.
Dupo uses Answer Set Programming (ASP)~\cite{gebser:asp2011} with the Clingo solver~\cite{gebser:clingo2014} to search the design suggestions, and translates the resulting ASP models (set of transformations) to Cicero rules.
Dupo's interface and recommender are backed by Svelte~\cite{svelte} and FastAPI~\cite{fastapi}.

\figureStudyOutcome{}

\section{User Study}
\noindent Motivated by our design considerations (\autoref{sec:guidelines}), we evaluated Dupo with six expert responsive visualization authors to answer the following research questions:
\emph{How well can authors explore alternative design ideas in Dupo (RQ1)?} 
\emph{How do authors combine manual edits and~automated design suggestions (RQ2)?}
\emph{How do authors use Dupo's recommender control options while interacting with the \menu{action widget} (RQ3)?}
We also consider general feasibility and usefulness questions:
\emph{Are Dupo's design suggestions and final outcomes deemed reasonable by authors (RQ4)?}
\emph{Do authors see Dupo as a feasible tool for their day-to-day responsive visualization tasks (RQ5)?}
\emph{What do authors see as the remaining challenges to address to improve Dupo's usefulness (RQ6)?}
For greater ecological validity, our study asked participants to create responsive versions for prior visualizations they had created.
Evaluating Dupo with participants' own use cases was also useful for helping us identify small improvements to Dupo to better support certain design situations while keeping the overall two-stage recommendation approach and other key features consistent across participants. 

\subsection{Methods}

\bstartf{Participants and Use Cases} We recruited six graphics reporters (E1--E6) who regularly published responsive visualizations on media outlets, snowballing via social media (Twitter, Mastodon) and Slack (News Nerdery, Data Visualization Society).
We asked participants to provide candidate use cases from their own work. 
We chose one of these examples that fit the scope of Dupo (in terms of supported visual encodings and interactivity) and exhibited variations in visual encoding combinations and layout. 
As shown in \autoref{fig:study_outcome}, participants contributed a map (E1), small multiples of line charts (E2), a layered and annotated line chart (E3), a stacked bar chart (E4), a 2D dot plot with many annotations (E5), and a grouped bar chart (E6).
For interactivity that Dupo does not support (E3: parallax; E4: view toggling), we took a single representative view (E3: conclusion chart; E4: default view). 

\bpstart{Study Procedure} 
Before each session, we asked two background questions about the design process behind each participant's contributed visualization and their preferred authoring tools.
During each remote-control Zoom session, the participant first went through a step-by-step guided training (creating a bar chart for desktop and exploring design suggestions for a phone) to get familiar with Dupo.
Then, participants spent 30 minutes completing the main study task: to create responsive versions for their desktop design (which we replicated in Dupo) while thinking aloud. 
We asked them to create the phone version first and then work on tablet, thumbnail, and print versions if time allowed.
While Dupo supports different directions of responsive design (\eg~mobile-first, simultaneous editing), we decided to fix the desktop-first direction for two main reasons: (1)~all participants reported desktop-first as their typical design direction when asked prior to the study, and (2)~our goal was to observe how they use Dupo's design suggestions along with manual editing, rather than how Dupo affects the direction of creation. 

After the main task, participants self-evaluated their work and then completed a 15-minute interview.
For self-evaluation, participants rated their satisfaction on a scale from 1 (Very unsatisfied) to 5 (Very satisfied): \textit{Given the limited design time you had, how satisfied are you with this draft?}
We followed up the self-evaluation by asking about any further edits they had wanted to make if given more time.
In the structured interview, we asked participants' about their overall reactions to seeing the design suggestions, comparison between Dupo and their day-to-day tasks, and Dupo's features that they would want for their usual tools. 
We then asked about the usefulness of the \pipe{Exploration} suggestions for the phone version by showing them the suggestions again.
Lastly, we asked about suggestions for general system improvement.
Each participant was compensated with USD 60 (gift card).
Detailed study protocols are included in the Supplementary Material.

\subsection{Results}
We analyzed the think-aloud and interview transcripts using top-down thematic analysis~\cite{braun2006:thematic} given the research questions outlined above.
We analyzed the think-aloud transcripts, system logs, and screen recordings together to identify design exploration patterns.
We find that Dupo supports exploring a range of responsive transformations and creating satisfactory designs, yet our participants suggested improvement for direct manipulations and different ways to show \pipe{Alteration} suggestions.

\bstart{Support for Design Exploration (RQ1)}
During the study sessions, participants made two to four responsive versions across different screen sizes that they were pleased with over a relatively short amount of time (about 30min), thereby demonstrating that Dupo supports creating multiple outcomes.
We observed participants reasoning about different design suggestions in various ways.
We tagged the task recordings with observations of how participants appeared to be reasoning about the designs as they interacted with the \menu{recommender window}.
First, we observed that participants reviewed \textit{most of the suggestions}, rather than just one or two that initially appealed to them, suggesting that the recommendation sets were compelling.

Second, participants frequently compared two or more suggestions, explicitly reflecting on which one was better for their purposes. 
For example, comparing suggestions for a stacked bar chart with small multiples for a tablet version, E4 noted that \textit{``[The multiple bar chart]'s a different look. It's an emphasis on the [regions] as opposed to the overall [volume] of the [independent book stores] like you get here [with the stacked bar chart].''} E4 then chose the small multiples (\autoref{fig:study_outcome} \logB{M4}, Tablet).
Participants also confirmed the need for authoring agency over small changes by considering possible manual edits to the suggestions, for example, by asking \textit{``Is it possible to manually move the annotation?''} (E1; \logB{M1}, Phone) or by saying \textit{``I'm gonna take out all the annotations for $\dots$ social [media]''} (E5; \logB{M5}, Social).

Overall, participants seemed relatively confident in the design choices they made, which they articulated by summarizing their decisions.
For instance, E3 first summarized that a slope chart suggestion for his thumbnail version \textit{``is only showing two values''} and then said,  \textit{``let's pick this one [with more detail]''} because he had already noted that he wanted to keep the details (\logB{M3}, Thumbnail).

In terms of fixation on well-known designs, participants in general made a phone version that seemed similar to their original phone versions.
This fixation is not necessarily surprising given that they had already gone through rigorous design iterations in a newsroom-like setting.
E2 mentioned, \textit{``It's hard to disconnect from my original decisions I made.''}
However, participants considered designs with more drastic changes for the versions that they had previously paid less attention to (\eg~tablet, thumbnail).
For example, given a small multiple line chart, E2 overlaid those lines for a thumbnail (\autoref{fig:study_outcome}~\logB{M2}). 
With a grouped bar chart for desktop, E6 chose a heatmap for a thumbnail because it seemed to \textit{``make people curious''} about the content, though she still wanted further aesthetic changes. 
We observed design suggestions for a previous version sometimes influencing participants' designs for later versions.
For example, E4 manually transposed the axes for the thumbnail~\logB{M4}, inspired by similar \pipe{Exploration} suggestions for the phone version.
Likewise, E5 removed annotations for the social version~\logB{M5}, pointing out \pipe{Alteration} (\menu{see similar}) suggestions for the phone version.

\bstart{Mixed-initiative Usage Patterns (RQ2)}
Given that participants immediately saw design suggestions upon creating a new responsive artboard, they tended to choose a design suggestion first before making manual edits.
Participants generally spent less time using the recommender ($m$=3.0min, $s$=1.9) than making manual edits ($m$=5.3, $s$=5.8) per version.
Yet, they took relatively less time to make minimal or already brainstormed manual edits, such as E2's column adjustment for small multiples (\autoref{fig:study_outcome} \logB{M2}, Tablet) and E5's social version inspired by the suggestions for the phone (\logB{M5}, Social).

Dupo was used to automate changes to overall sizes (scaling the chart size, text size, \etc), layout (transposing), and encoding, while participants tended to manually edit text elements (annotations and titles) and reference objects, which are essential in narrative visualization.
For example, E4 changed the title text for text alignment and justification. 
After making an edit, participants tended to closely examine every element to ensure that all important information was preserved, which helped them identify what they might need to edit next.
After moving annotations in the phone version, for example, E5 realized that he needed to further adjust the spacing.
E6 requested recommendations after making some manual edits, and then branched an \pipe{Alteration} suggestion.
Then, E6 made more edits to the two candidate artboards simultaneously via edit propagation, before finalizing one of them.

Participants tended to make manual edits by testing out a variety of different values for the visual encoding (\eg~by repeatedly changing the \textit{x} or \textit{y} value to position an annotation appropriately).
When we filtered out manual edits (written as Cicero rules) dealing with the same element (\ie~the same specifier and option properties) from the log of participants' manual edits,
the average number of manual edits reduced from 85.3~($s$=82.1) to 34.8~($s$=20.5).
In contrast, participants applied an average of 28.3~($s$=12.7) edits by the recommender.
We note that each recommender rule changed multiple properties, and participants explored multiple design suggestions with at most fifteen rules. 
Overall, Dupo enabled participants to focus on edits that needed careful visual inspection, while expediting high-level responsive transformations.

\bstart{Users' Control and Interpretability (RQ3)}
Participants interacted with features in the \menu{recommender window}, and often read the descriptions provided in the \menu{action widget}.
For example, E6 actively used the \menu{hide this} features, and she valued \menu{hide this} because \textit{``it removes all the noise that you don’t need, and you can focus on the limited options that you will get.''} 
Using the \menu{bring my existing edits} feature, E6 said, \textit{``[it] is a learning experience $\dots$ in terms of how I can do better [because I could] compare what I did earlier [manually] and some new suggestions.''}
Meanwhile, E4 wished for more control in selecting a source design by asking, \textit{``Can I make a thumbnail based on the mobile version?''} instead of the desktop version.
E4 also wanted to use the \menu{hide this} feature independently for each version. 

\bstart{Design Satisfaction (RQ4)}
Across all sessions, participants created and rated their satisfaction for 17 designs.
In general, participants were satisfied (seven ``very satisfied'' and seven ``satisfied'') with the versions they created. 
Some responded ``just okay'' to their last versions because they did not have enough time to complete an initial version (\eg~E3, Print) or they wished to fine-tune the designs (\eg~E2, E4, Thumbnail).

Participants in general found our design suggestions to be reasonable and realistic potential improvements to their design process. 
E4 said, \textit{``[Dupo] literally made 90\% of the chart I would have made by hand on my own for a mobile [phone].''}
E2 mentioned that his team \textit{``actually did mock up a version that was like this [heatmap].''}
Even though participants sometimes thought some design suggestions were less visually appealing due to drastic changes from the desktop versions or less preferred label arrangement, all participants seemed able to understand why those examples were suggested. 
For example, E3 did not like transposing the line chart, yet he acknowledged that \textit{``in some cases, it can also be helpful.''}
E6 said some axis-transpose options were not aligned with her intention with the original bar grouping.

When we asked participants to rank the \pipe{Exploration} suggestions for the phone version during the follow-up interview, they exhibited several criteria, such as similarity with the source design and well-formedness.
For example, E2 ranked the two suggestions with encoding changes (to a heatmap) in the third and fifth place, saying that they were \textit{``missing the point [of] $\ldots$ what the desktop one was doing.''}
Considering the well-formedness of the design, E6 said, \textit{``Labels are aligned better and I can see all the values''} as she ranked the design suggestion she chose for the phone version~\logB{M6} in first place. 
Participants found some \pipe{Exploration} suggestions to be subtle in a way that we did not expect previously.
While comparing aggregated and non-aggregated \pipe{Exploration} alternatives, for example, E1 was not able to figure out the difference at first.
Once we clarified the aggregation, E1 ranked the aggregated version in the first place, saying, \textit{``that's actually really~smart.''}

\bstart{Feasibility and Remaining Challenges (RQ5 \& RQ6)}
Comparing to their day-to-day tools, participants believed Dupo could help reduce the time required to prototype design ideas by providing high-fidelity designs.
For example, E1 noted that \textit{``I [had to] create three [responsive] versions. It takes up a lot of time. Even though you can copy and paste, there's still a lot of manual work going on.'' }
E4 said that, \textit{``It wasn't just a little pencil sketch or a crappy chart that needed work. Basically, it did all that work.''}
Pointing out that \textit{``mobile is like a whole different world to think about,''} E2 said Dupo could help to \textit{``get into that mindset much easier.''}
E6 mentioned that being able to have different designs for responsive versions, which her tool (Flourish) did not support, would allow her team to consider further encodings like treemaps for desktop versions with simplified mobile versions.

However, those familiar with Adobe Illustrator or similar tools pointed out the needs for extended direct manipulation. 
For example, E5’s visualization was annotation-heavy, so positioning the annotation and adjusting the spacing accordingly were important tasks. 
E5 found it inconvenient to make these changes in Dupo due to the limited direct manipulation. 
Participants also needed additional support for inspecting \pipe{Alteration} suggestions as they often included more subtle changes.
E3 proposed suggesting ``edits'' rather than ``designs,'' and letting users test out different edits to help them make sense of what is being changed. 
\section{Discussion}
We implemented Dupo, a mixed-initiative tool for creating responsive visualizations, to support design exploration, users' agency over recommendation procedures, progressive authoring, and flexible artboard management.
Dupo provides two-step design suggestions (\pipe{Exploration} and \pipe{Alteration}), edit \pipe{Augmentation} (\menu{quick edits}), various user control options (\eg~the \menu{action widget}), and edit propagation across artboards as well as between manual and automated designs.
We evaluated Dupo with six expert data journalists using their own visualization cases, observing that Dupo supports exploring different design ideas, helping users to come up with satisfactory outcomes.
Below, we discuss implications to other responsive visualization design approaches and future challenges for mixed-initiative authoring with adaptive systems.

\subsection{Extension to Other Responsive Visualization Methods}
\noindent Dupo employs a graphic-based interface, similar to Adobe Illustrator and Canva, that allows for visually manipulating chart elements; this approach has several benefits with respect to implementing the recommender. 
First, our approach can regularize the representation of user input for the recommender and hence easily specify the roles of chart elements, such as using our extended Vega-Lite~\cite{satyanarayan:vega-lite2017} for the visualizations and Cicero~\cite{kim:cicero2022} for responsive transformations.
In contrast, users can have more freedom in representing their designs with programming-based approaches (\eg~writing code for D3.js~\cite{bostock:d32011}), which poses more complexity in interpreting different visualization roles.
Second, Dupo can visually demonstrate design suggestions within its graphical interface, whereas programming-based approaches may require additional rendering processes.
Third, Dupo supports flexibility in customizing responsive artboards without necessarily maintaining the same mark type, encoding channels, or text elements, compared to the template-based tools (\eg~Datawrapper~\cite{datawrapper} or Flourish~\cite{flourish}), as pointed out by~E6.

Our user study participants generally leverage a diverse set of visualization tools including template-based, programming-based, and graphic-based tools.
To better support different authoring experiences, it is important to extend our mixed-initiative approach to different types of tools. 
As noted above, however, there are several challenges to do so for template-based and programming-based tools, illuminating future research opportunities. 
Using a template-based tool, for example, users choose a template of fixed responsive designs, and they can only use those designs.
Instead, such tools could allow for combining each responsive version from different templates, so that a recommender can search designs more flexibly and users can have more freedom with their designs, as noted by E2 and E6.
Doing so would require asking research questions like how to search over a template design space in a way that supports preserving design information across responsive versions.
For programming-based approaches, recommenders could employ an extraction model~\cite{d3restyle:harper2014} or AI code suggestion~\cite{copilot} for responsive visualization.
Such approaches will benefit from real-time bundlers (\eg~Rollup~\cite{rollup}) that renders the outcome as a user makes changes to the code, potentially enabling immediate visual demonstration of design recommendations expressed programmatically.

\subsection{Next Step: Adaptive Recommendation}
Compared to analytic domains like exploratory data analysis, responsive design involves a high degree of creativity in order to craft and effectively communicate a narrative across different end-user devices.
Many organizations that require responsive design may develop their own individual preferences or design patterns that are necessary to incorporate into the design process.
In addition, we will discover new design patterns for responsive visualization as the area is evolving, demanding an extended search space.
To make this process easier, responsive visualization tools can employ an adaptive recommender that updates its search space by learning design patterns from users. 
While Dupo offers implicit user constraints using \menu{hide-this}, adaptive mixed-initiative systems can learn and share new responsive techniques across different users.
This approach opens up several challenges for future work in addition to developing an adaptive mechanism.
For example, such a system may need to determine whether to learn a design pattern of a user and whether one user's technique would be useful for other users.
It is also necessary to ask how to support users to easily control the outcome of the recommenders with a growing search space.

\subsection{Limitations and Future Work}
Dupo currently considers common mark types, encodings, and interactions supported by Vega-Lite~\cite{satyanarayan:vega-lite2017}. 
While we outlined high-level rationales for generating \pipe{Exploration} design suggestions, we do not claim that our search space is exhaustive. 
Thus, future work could extend Dupo with regard to encodings and design suggestions.
Next, our user study only considered the desktop-first approach.
A longer term user study could observe how mixed-initiative approaches impact responsive visualization authoring in different settings like collaborative creation or simultaneous editing and improve the system with adaptive learning.
Lastly, we qualitatively evaluated Dupo to elicit perspectives of expert users.
Future work could compare future responsive visualization tools with Dupo as a benchmark.

\section{Conclusion}
We presented Dupo, a mixed-initiative responsive visualization authoring tool, that augments users' design exploration with recommendations to support authoring multiple artboards while retaining user control.
Our qualitative evaluation with six expert users on their pre-existing, real-world responsive use cases shows that Dupo's two-step recommendation pipeline can support exploring responsive designs by reducing redundancy and tedious work required for prototyping alternatives.

\acknowledgments{%
Special thanks to Eunyee Koh and Priyanka Nanayakkara for their valuable feedback and support. 
}

\vspace{-1mm}

\section*{Supplementary Material}
In the Supplementary Material (\url{https://see-mike-out.github.io/dupo-supplementary}), we provide (1)~details for our pilot study, (2)~the Dupo system documentation (user manual), (3)~a video demo of Dupo, (4)~a walkthrough video of using Dupo, (5)~high-level rationales for each \pipe{Exploration} suggestion, (6)~details for the ranking measures, (7)~the main study protocols, and (8)~the study outcomes (visualization designs and suggestions). 
A public demo of Dupo is available at \url{https://see-mike-out.github.io/dupo-public}.

\vspace{-1mm}

\bibliographystyle{abbrv-doi-hyperref}

\bibliography{bibliography}


\appendix 

\end{document}